\documentclass[twocolumn,appendixfloats]{aastex6}
\usepackage{graphicx}
\usepackage{natbib}
\usepackage{latexsym}
\usepackage{amssymb}
\usepackage{longtable}
\usepackage{amsmath}
\usepackage{url}
\def\msun{\ifmmode {\rm\,M_\odot}\else ${\rm\,M_\odot}$\fi}
\def\Msun{\ifmmode {\rm\,\it{M_\odot}}\else ${\rm\,M_\odot}$\fi}
\def\lsun{\ifmmode {\rm\,L_\odot}\else ${\rm\,L_\odot}$\fi}
\def\Lsun{\ifmmode {\rm\,\it{L_\odot}}\else ${\rm\,L_\odot}$\fi}
\def\rsun{\ifmmode {\rm\,R_\odot}\else ${\rm\,R_\odot}$\fi}
\def\Rsun{\ifmmode {\rm\,\it{R_\odot}}\else ${\rm\,R_\odot}$\fi}
\def\Tsun{\ifmmode {\rm\,T_\odot}\else ${\rm\,T_\odot}$\fi}
\def\arcsec{\ifmmode {^{\prime\prime}}\else $^{\prime\prime}$\fi}
\def\asec{\ifmmode {^{\prime\prime}}\else $^{\prime\prime}$\fi}
\def\arcmin{\ifmmode {^{\prime}}\else $^{\prime}$\fi}
\def\amin{\ifmmode {^{\prime}}\else $^{\prime}$\fi}
\def\simlt{\mathrel{\spose{\lower 3pt\hbox{$\mathchar"218$}}\raise 2.0pt\hbox{$\mathchar"13C$}}}
\def\simgt{\mathrel{\spose{\lower 3pt\hbox{$\mathchar"218$}}\raise 2.0pt\hbox{$\mathchar"13E$}}}

\begin{document}

\title{A search for H$\alpha$ absorption around KELT-3 b and GJ 436 b}

\author{P. Wilson Cauley and Seth Redfield}
\email{pcauley@wesleyan.edu}
\affil{Wesleyan University\\
Astronomy Department, Van Vleck Observatory, 96 Foss Hill Dr., Middletown, CT 06459}
\affil{Visiting astronomer, Kitt Peak National Observatory, National Optical Astronomy Observatory, which is operated by the Association of Universities for Research in Astronomy (AURA) under a cooperative agreement with the National Science Foundation.}

\author{Adam G. Jensen}
\affil{University of Nebraska-Kearney\\
Department of Physics \& Astronomy, 24011 11th Avenue, Kearney, NE 68849}

\begin{abstract} 

Observations of extended atmospheres around hot planets have generated exciting results concerning the
dynamics of escaping planetary material. The configuration of the escaping planetary gas can result in
asymmetric transit features, producing both pre- and post-transit absorption in specific atomic transitions.
Measuring the velocity and strength of the absorption can provide constraints on the mass loss mechanism
and, potentially, clues to the interactions between the planet and the host star. Here we present a search 
for H$\alpha$ absorption in the circumplanetary environments of the hot planets KELT-3 b and GJ 436 b. We
find no evidence for absorption around either planet at any point during the two separate transit epochs that
each system was observed. We provide upper limits on the radial extent and density of the excited hydrogen
atmospheres around both planets. The null detection for GJ 436 b contrasts with the strong Ly$\alpha$ 
absorption measured for the same system, suggesting that the large cloud of neutral hydrogen is almost
entirely in the ground state. The only confirmed exoplanetary H$\alpha$ absorption to date has been made
around the active star HD 189733 b. KELT-3 and GJ 436 are less active than HD 189733, hinting that
exoplanet atmospheres exposed to EUV photons from active stars are better suited for H$\alpha$
absorption detection.

\end{abstract}

\keywords{}

\section{INTRODUCTION}
\label{sec:intro}

Due to the large amount of stellar flux received, hot planets ($P_{orb}$ $\lesssim$ 5 days) provide insight into extreme 
astrophysical processes that do not occur around longer period planets. These planets are often observed to be
inflated \citep[e.g.,][]{laughlin11} and detailed rotational and atmospheric dynamics have now been observed for
the hot Jupiters HD 189733 b \citep{wyttenbach,brogi16,louden} and HD 209458 b \citep{snellen10}. 

Perhaps the most interesting dynamical process observed around hot planets is evaporative mass loss.
This process was first observed by \citet{vidal} for the hot Jupiter HD 209458 b: observations of
the UV hydrogen line Ly$\alpha$ showed an extended atmosphere of neutral hydrogen that was larger
than the planet's Roche limit.  Followup observations demonstrated that the mass loss was highly
variable and changed significantly from epoch to epoch \citep{desetangs12}. Subsequent
investigations of other hot planet systems, namely HD 189733 b, 55 Cnc b, and GJ 436 b, have
revealed a handful of mass loss detections \citep{desetangs,ehren12,kulow,ehren15}.

The most recent mass loss detection around GJ 436 b by \citet{ehren15} highlighted the possibility
of observing these extended and escaping atmospheres both before and after the nominal planetary
transit (i.e., the broadband white light transit). Besides GJ 436 b, evidence for pre-transit
signatures have been observed for WASP-12 b \citep{fossati} and HD 189733 b
\citep{benjaffel,bourrier13,cauley15,cauley16}.

Although it seems that both pre- and in-transit signatures of extended atmospheres may be common for
hot planets, observing these phenomena requires significant telescope resources since there is
evidence that they do not manifest in broadband photometric observations \citep{turner16}, i.e., the
absorption is only seen in strong atomic lines. Due to the abundance of hydrogen and the intrinsic
line strength, the atomic transition most suited for measuring the dynamics of the escaping material
is the resonance line Ly$\alpha$ \citep[e.g.,][]{desetangs,ehren15}.  Currently, the only telescope
capable of performing high-spectral resolution Ly$\alpha$ observations is the \textit{Hubble Space
Telescope} (\textit{HST}). Thus it is difficult to perform exploratory Ly$\alpha$ measurements of
these phenomena for a large number of systems due to the high demand for time on \textit{HST} and
the inherent variability associated with the mass loss process \citep[e.g.,][]{desetangs12}.  

Our recent detections of pre-transit, as well as in-transit, H$\alpha$ absorption signatures around
HD 189733 b suggest that it may be possible to detect highly extended neutral atmospheres using
high-resolution optical spectrographs \citep[][]{jensen12,cauley15,cauley16}. A detection of excited
hydrogen via the flux decrement at the Balmer jump was first reported by \citet{ballester07} for HD
209458 b.  While our excited hydrogen measurements for HD 189733 b do not show evidence of large
blue-shifted velocities indicative of escaping material, the transit depths indicate that the
transmission spectrum is probing pressures of 10$^{-6}$--10$^{-9}$ bars, i.e., the planetary
thermosphere \citep{christie}. We note that \citet{barnes16} recently questioned the
planetary origin of the HD 189733 H$\alpha$ signal due to velocity centroids in the transmission
spectra that appear to correspond to the stellar rest frame. Thus there are still questions
concerning the nature of the in-transit HD 189733 H$\alpha$ measurements.  In an upcoming paper,
however, we will argue for the planetary interpretation although further observations are certainly
warranted to confirm either hypothesis. Detections of highly extended atmospheres via H$\alpha$
absorption would provide strong targets for followup Ly$\alpha$ observations that can be used to
constrain the mass loss mechanism and escape rate.

In this paper we present high spectral resolution observations of H$\alpha$ for the hot planets
KELT-3 b \citep{pepper13} and GJ 436 b \citep{butler04} with the aim of searching for H$\alpha$
absorption in the circumplanetary environment.\footnote{We include the bound extended atmosphere in
this phrase.} KELT-3 b was chosen for both the brightness (V$=$9.8), visibility during the 2016 A
semester, and the suggested low chromospheric activity level of its host star \citep{pepper13}, a
desirable property since stellar activity can mimic absorption signatures
\citep[e.g.,][]{berta,cauley16}. GJ 436 b was targeted due to the large Ly$\alpha$ transit depth
measured by \citet{ehren15}: even a small fraction ($\sim$0.01\%) of the neutral hydrogen in the
$n=2$ state would be detectable across a single transit. A detection of excited hydrogen absorption
around these planets would significantly expand the confirmed H$\alpha$ detections
\citep[see][]{jensen12,cauley15,cauley16} and further our understanding of how this line forms in
hot exoplanet environments.

\section{OBSERVATIONS AND DATA REDUCTION}
\label{sec:observations}

The observations were performed using the Bench Spectrograph with Hydra \citep{barden92} at the WIYN 3.5 meter telescope. 
The instrument was configured to provide the highest resolution possible ($R\sim20,000$) for a single spectral order
of width 380 \AA\ with H$\alpha$ positioned at the center of the order. This setup was achieved using the Bench Spectrograph Camera
combined with the blue fiber cables and the 316@63.4 echelle grating. Target exposures were taken for two separate 
transits of each object. Details of the observations are given in \autoref{tab:tab1}. We note that the smaller number of in-transit 
exposures of KELT-3 on 2016-02-04 compared with 2016-03-02 is the result of suspended observations due to high wind speeds.

\begin{deluxetable*}{lccccccccccc}
\tablecaption{Log of observations\label{tab:tab1}}
\tablehead{\colhead{}&\colhead{$V$}&\colhead{Date}&\colhead{$t_{start}-t_{mid}^a$}&\colhead{$t_{end}-t_{mid}$}&\colhead{$t_{exp}$}&\colhead{}&\colhead{}&\colhead{}&
\colhead{$\overline{S/N}$}&\colhead{}&\colhead{}\\
\colhead{Object}&\colhead{(mag)}&\colhead{(UT)}&\colhead{(hours)}&\colhead{(hours)}&\colhead{$(seconds)$}&\colhead{$N_{pre}$}&\colhead{$N_{in}$}&\colhead{$N_{post}$}&\colhead{(near 6500 \AA)}&\colhead{$N_{comp}$}&\colhead{$N_{sky}$}\\
\colhead{(1)}&\colhead{(2)}&\colhead{(3)}&\colhead{(4)}&\colhead{(5)}&\colhead{(6)}&\colhead{(7)}&\colhead{(8)}&\colhead{(9)}&\colhead{(10)}&\colhead{(11)}&\colhead{(12)}}
\tabletypesize{\scriptsize}
\startdata
KELT-3 & 9.8 & 2016-02-04 & $-$4.4 & 4.6 & 600/900 & 16 & 7 & 11 & 183 & 11 & 64 \\
             &  & 2016-03-02 & $-$6.4 & 2.2 & 600 & 27 & 16 & 3 & 190 & 11 & 60 \\
GJ 436 & 10.7 & 2016-02-15 & $-$3.0 & 2.0 & 1500 & 6 & 2 & 4 & 133 & 10 & 62 \\
             &  & 2016-02-23 & $-$4.4 & 3.3 & 1200 & 12 & 2 & 9 & 142 & 10 & 59 \\
\enddata
\tablenotetext{a}{$t_{mid}$ denotes the mid-transit time.}
\end{deluxetable*}

A single fiber is dedicated to the target. The remaining fibers are either assigned to a sky position or to a comparison
star in the field. The exact number of sky and comparison object fibers for each date are given in \autoref{tab:tab1}.
The sky fibers are necessary to subtract out a simultaneous sky spectrum, which includes both continuum
emission and night sky emission lines. Comparison stars are important for identifying artifacts in the transmission spectrum,
e.g., imperfect sky subtraction.

\subsection{Data reduction}
\label{sec:datareduction}

The data were reduced using custom IDL routines. Standard reduction steps were performed including bias
subtraction, flat fielding of the individual fiber spectra, and wavelength calibration using ThAr lamp spectra. Individual
fiber spectra were extracted using a 7-pixel wide boxcar. Cosmic rays in all spectra were identified and removed post-extraction
through median filtering of the individual exposures.   

The background sky emission can be a non-negligible contribution to the object spectrum depending on the
relative brightness between the sky and the object. This contribution can be exacerbated by moonlight scattered
off of clouds in the field of view. Thus a careful accounting of the sky spectrum from exposure to exposure is important since we are 
looking for signals in the target spectrum at the $\sim$1\% level. We also account for small variations in the
sky spectrum across the field of view by only selecting the sky fibers with angular distance $\leqslant$10$'$ from the
target or comparison fibers. This normally includes $\sim$10--15 sky fibers for each object fiber. 

\begin{figure*}[h]
   \centering
   \includegraphics[scale=.73,clip,trim=12mm 35mm 8mm 100mm,angle=0]{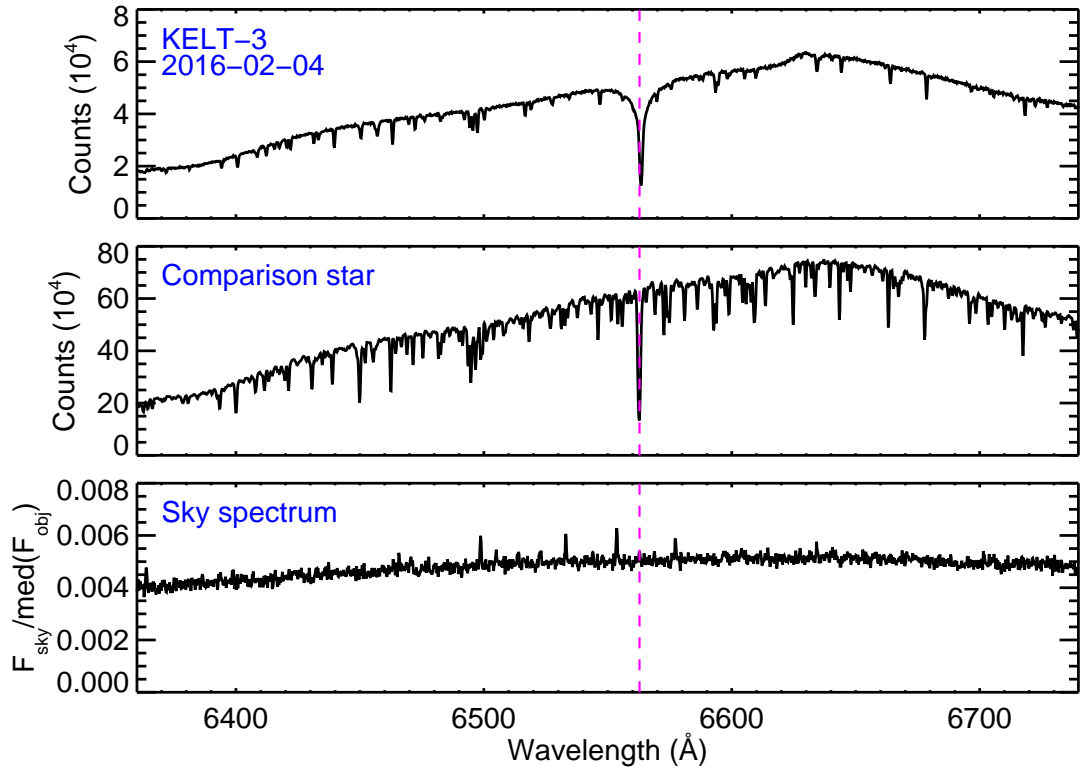} 
   \figcaption{Raw extracted spectra for a single exposure of KELT-3 from 2016-02-04. The sky spectrum is fairly free of strong
   emission lines and is $\sim$0.4\% of the object signal.\label{fig:kelt3raw}}
\end{figure*}

\begin{figure*}[h]
   \centering
   \includegraphics[scale=.73,clip,trim=12mm 35mm 8mm 100mm,angle=0]{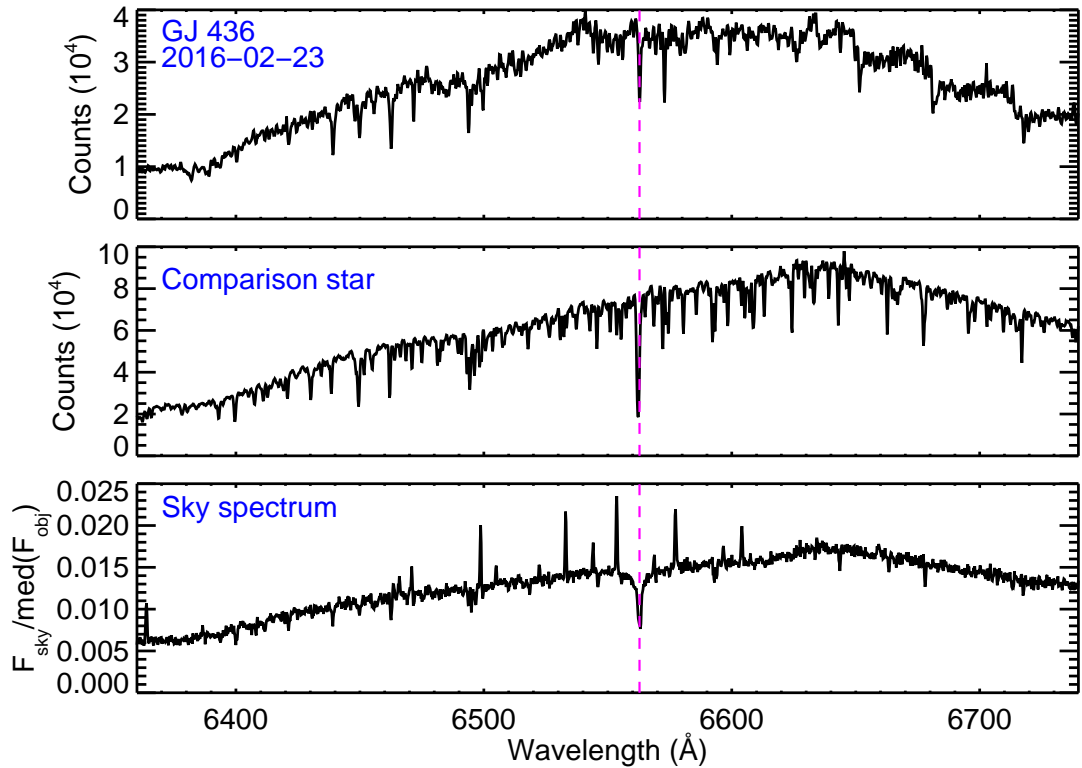} 
   \figcaption{Raw extracted spectra for a single exposure of GJ 436 from 2016-02-23. Note the strong H$\alpha$ absorption
   line in the sky spectrum. This is moonlight reflected from clouds in the field of view. The accumulated sky counts
   are a much larger percentage ($\sim$1.5\%) of the object spectrum than for the KELT-3 nights.\label{fig:gj436raw}}
\end{figure*}

For each exposure the proximate sky fibers are selected for both the target fiber and each individual comparison fiber. 
The relative transmission efficiency of the sky and target/comparison fibers are calculated using the extracted flat field spectra.
Cosmic rays are filtered out of the sky spectra using a 3$\sigma$ median filter. The median sky spectrum for the target and each
comparison fiber is calculated and subtracted from the target or comparison spectrum. Examples of the simultaneous
target, comparison, and sky spectra are shown in \autoref{fig:kelt3raw} and \autoref{fig:gj436raw}.

Small shifts of a spectrum on the detector can occur throughout the night which amount to small changes in the
wavelengths of spectral features. To account for these changes, each spectrum is shifted against a portion of a 
reference spectrum until the standard deviation of ratio of the spectra is minimized. The region of interest near H$\alpha$ 
is not included in the standard deviation minimization. These shifts are typically
on the order 0.001 \AA--0.01 \AA\ and correlate strongly between objects, i.e., the cause is instrumental and not
intrinsic to the targets.

\section{H$\alpha$ transmission spectra and absorption measurements}
\label{sec:tspec}

We define the transmission spectrum as:

\begin{equation}\label{eq:strans}
S_T=\frac{F_{t_1}}{F_{t_2}}-1
\end{equation}

\noindent where $F_{t_1}$ is the average spectrum from some time period (e.g., $t_1 = $the pre-transit exposures and 
$t_2 = $the in-transit exposures) and $F_{t_2}$ is the average spectrum from a comparison time period. 
This is the same definition used in our previous work \citep{jensen12,cauley15,cauley16}.

\begin{figure*}[h]
   \centering
   \includegraphics[scale=.6,clip,trim=0mm 15mm 5mm 40mm,angle=0]{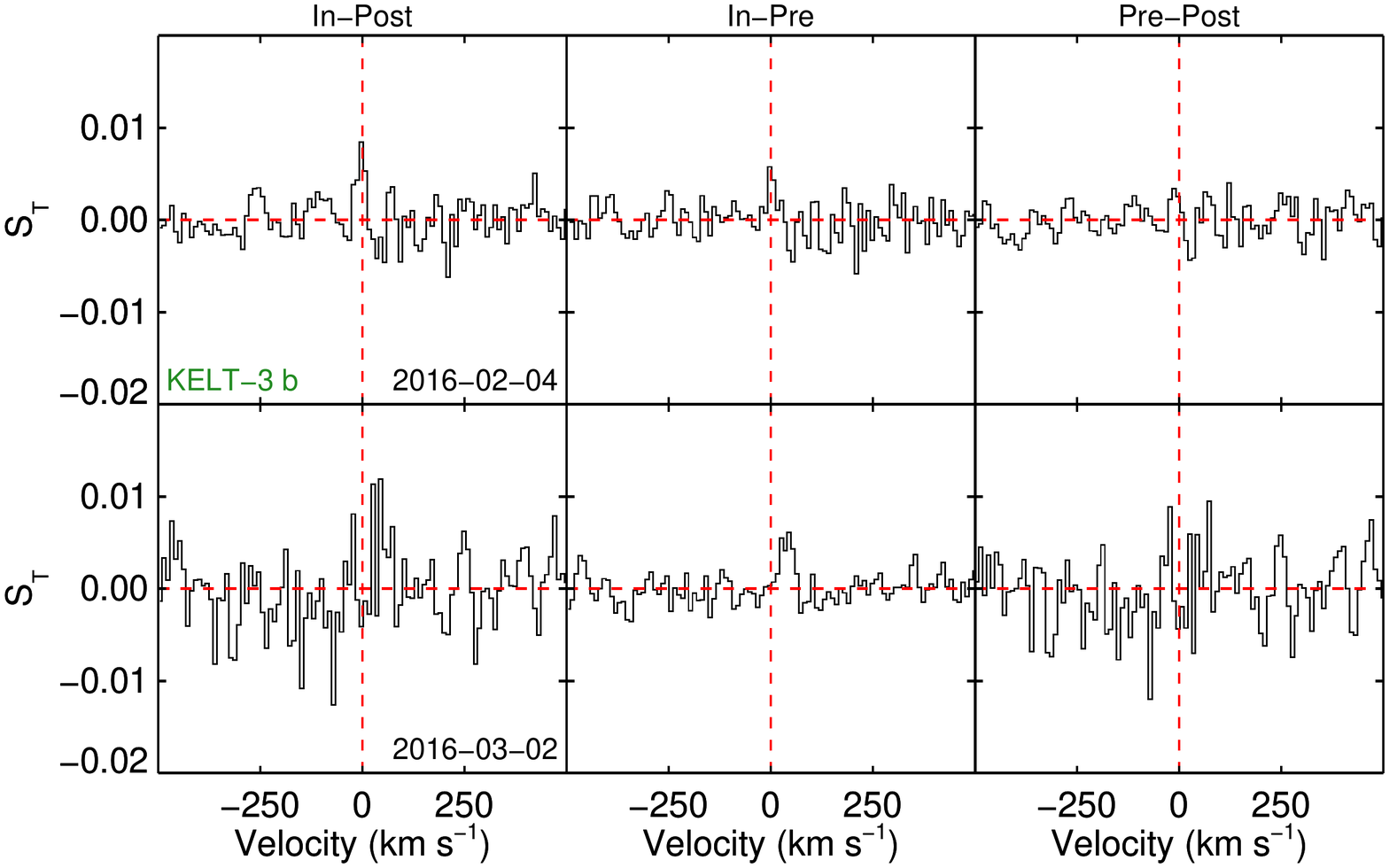} 
   \figcaption{Transmission spectra for KELT-3. The In-Post comparison is shown in the first column, In-Pre in the second, 
   and Pre-Post in the third. No significant absorption is detected in any of the comparisons. The measured values of
   $W_{H\alpha}$ and their uncertainties are given in \autoref{tab:tab2}.\label{fig:kelt3}}
\end{figure*}

\begin{figure*}[h]
   \centering
   \includegraphics[scale=.6,clip,trim=0mm 15mm 5mm 40mm,angle=0]{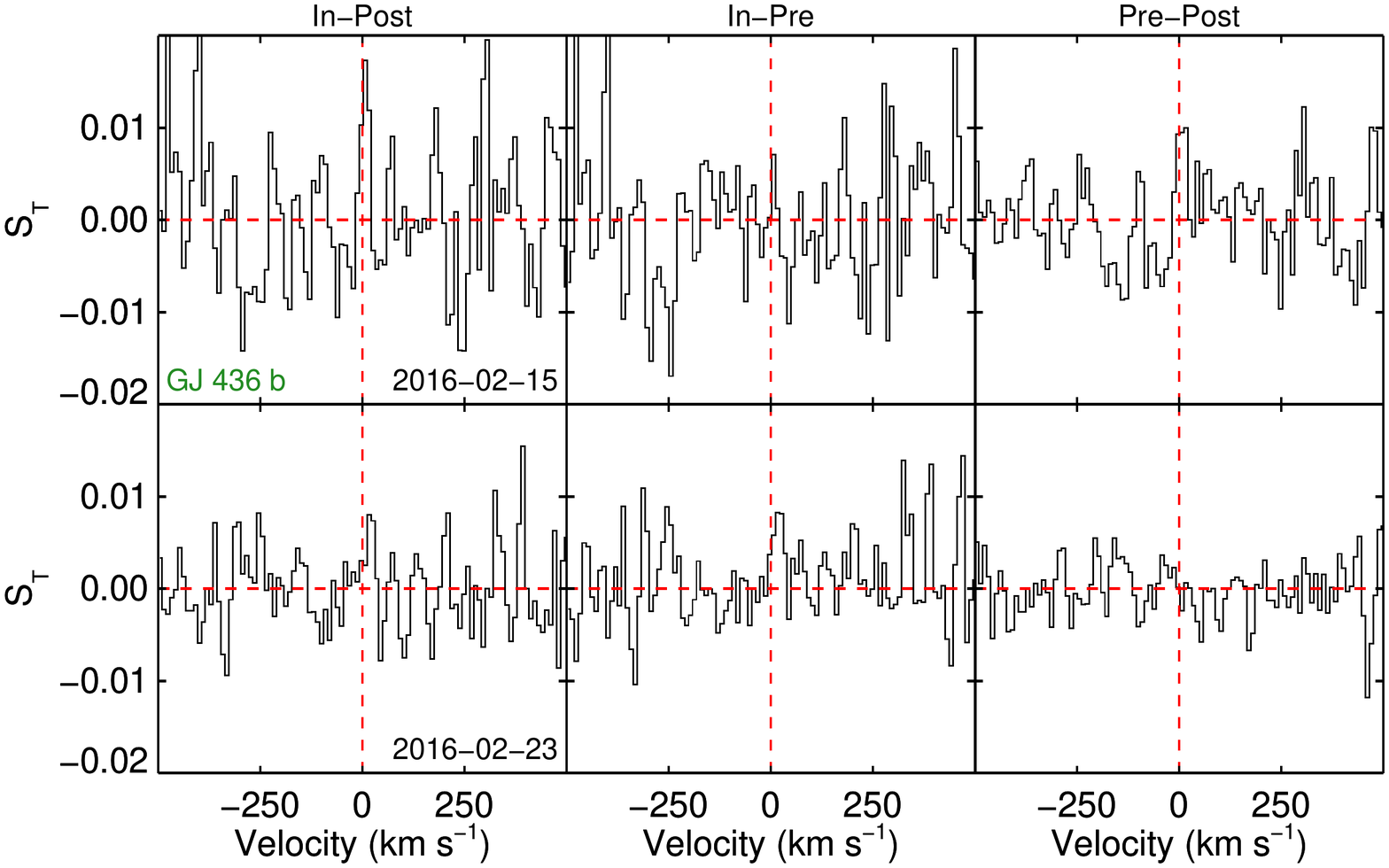} 
   \figcaption{Same as \autoref{fig:kelt3} except for GJ 436. The 2016-02-15 data are fairly low
   signal-to-noise and offer very weak constraints on an absorption signature. The 2016-02-23 data
   are higher signal-to-noise but no absorption is detected (see \autoref{tab:tab2}).\label{fig:gj436}}
\end{figure*}

We compare three different groups of exposures to each other: pre-transit, in-transit, and post-transit (see \autoref{tab:tab2}). 
After $S_T$ is calculated, it is renormalized with a low degree polynomial in order to remove any residual slope in the spectrum.
The transmission spectra for KELT-3 b are shown in \autoref{fig:kelt3} and the spectra for GJ 436 b are shown in \autoref{fig:gj436}.  
We measure the absorption across a 1000 km s$^{-1}$ band, or 21.9 \AA, centered on H$\alpha$. The absorption is calculated
as the equivalent width of $S_T$ from \autoref{eq:strans}:

\begin{equation}\label{eq:wlambda}
W_{H\alpha} = \sum\limits_{v=-500}^{+500} \left(1-\frac{F_{t_1}^v}{F_{t_2}^v} \right) \Delta\lambda_v
\end{equation}

\noindent where $F^v_{t_1}$ is the flux in the spectrum of the group of interest at velocity $v$, $F^v_{t_2}$ is the flux in the
comparison group spectrum at velocity $v$, and $\Delta\lambda_v$ is the wavelength difference at velocity
$v$. This is essentially the negative total of $S_T$. The units of $W_{H\alpha}$ are angstroms. The uncertainty
on $W_{H\alpha}$ is derived by summing the normalized flux errors in quadrature across the same integration
band.

\begin{deluxetable*}{lccccccccc}
\tablecaption{H$\alpha$ absorption measurements\label{tab:tab2}}
\tablehead{\colhead{}&\colhead{}&\multicolumn{2}{c}{In-Pre$^a$}&\colhead{}&\multicolumn{2}{c}{In-Post}&\colhead{}&\multicolumn{2}{c}{Pre-Post}\\
\colhead{Object}&\colhead{UT Date}&\colhead{$W_{H\alpha}$}&\colhead{$\sigma_{W_{H\alpha}}$}&\colhead{}&\colhead{$W_{H\alpha}$}&
\colhead{$\sigma_{W_{H\alpha}}$}&\colhead{}&\colhead{$W_{H\alpha}$}&\colhead{$\sigma_{W_{H\alpha}}$}\\
\colhead{(1)}&\colhead{(2)}&\colhead{(3)}&\colhead{(4)}&\colhead{}&\colhead{(5)}&\colhead{(6)}&\colhead{}&\colhead{(7)}&\colhead{(8)}}
\tabletypesize{\scriptsize}
\startdata
KELT-3 & 2016-02-04 & 1.1 & 4.0 && $-$0.3 & 4.5 && $-$3.5 & 3.7  \\
             & 2016-03-02 & 1.4 & 3.4 && $-$0.6 & 4.4 && $-$4.0 & 7.4  \\
GJ 436 & 2016-02-15 & $-$24.1 & 13.9 && $-$18.3 & 15.1 && $-$8.6 & 9.2 \\
             & 2016-02-23 & 9.5 & 9.1 && $-$0.4 & 8.7 && $-$0.7 & 6.0 \\
\enddata
\tablenotetext{a}{All $W_{H\alpha}$ and $\sigma_{W_{H\alpha}}$ values are in units of 10$^{-3}$ \AA.}
\end{deluxetable*}

Values of $W_{H\alpha}$ and its uncertainty are given in \autoref{tab:tab2}. There is no significant
absorption present in any of the comparisons.  The KELT-3 spectra are high enough signal-to-noise to
exclude 3$\sigma$ absorption $>$1.3$\times$10$^{-2}$ \AA, a value similar to that measured for HD
189733 b in \citet{cauley16}.

\section{Absorption timeseries}
\label{sec:timeseries}

\begin{figure*}[tbh]
   \centering
   \includegraphics[scale=.7,clip,trim=10mm 15mm 5mm 25mm,angle=0]{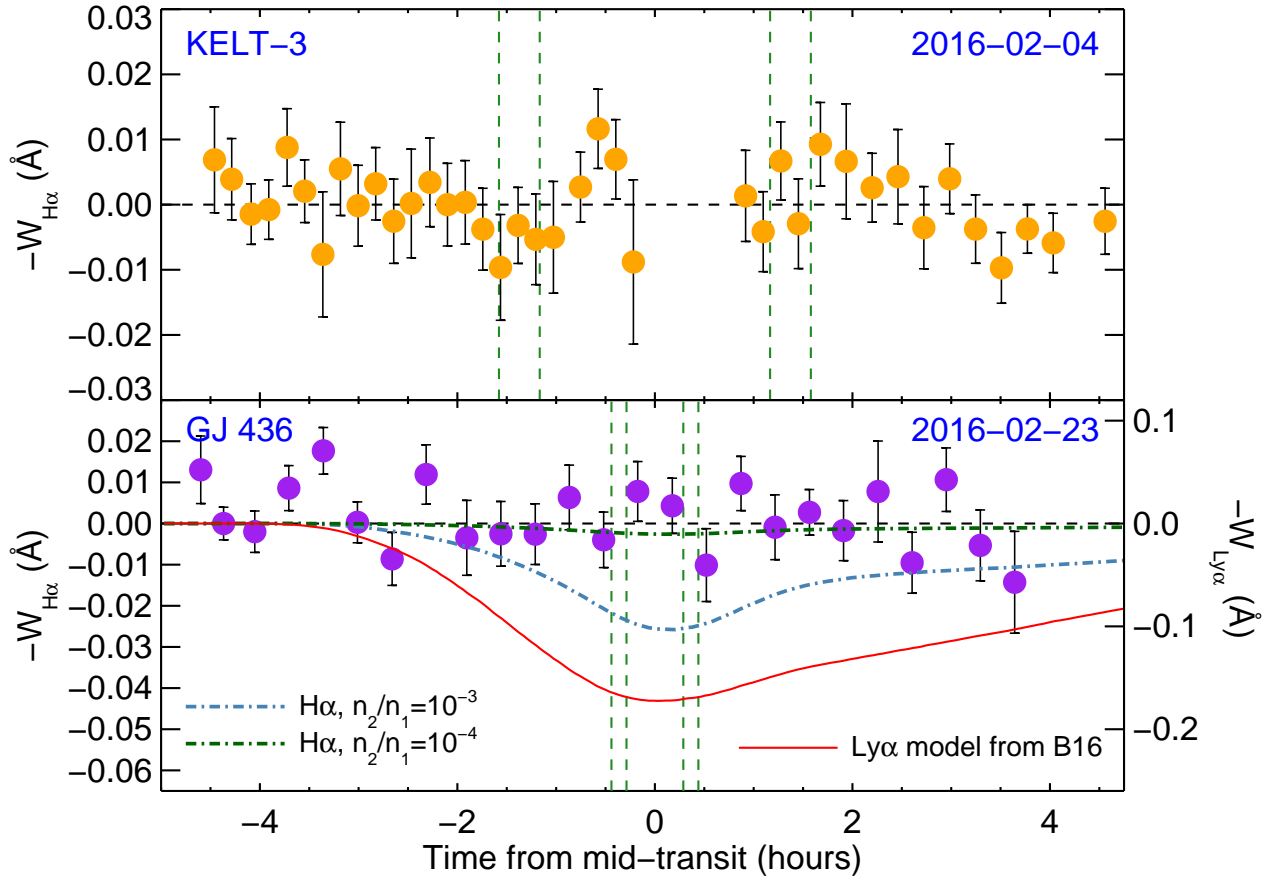} 
   \figcaption{$W_{H\alpha}$ timeseries for the higher signal-to-noise dates. Note the different scales for the upper panel
   (KELT-3) and lower panel (GJ 436). The vertical green dashed lines represent optical transit contact points. The red
   solid line in the lower panel shows the equivalent width measure $W_{Ly\alpha}$ calculated from the model presented in \citet{bourrier16} (B16).
   The dashed-dotted lines shows the simulated $W_{H\alpha}$ values assuming $n_2/n_1$ values of  
   10$^{-3}$ (steel blue) and 10$^{-4}$ (dark green). The right-hand axis gives the appropriate scale for the Ly$\alpha$ model curve. 
   The H$\alpha$ models show that our data are sensitive to excitation ratios of $\gtrsim$10$^{-3}$ for the 
   geometry presented in \citet{bourrier16}. Any ratio smaller than this results in $W_{H\alpha}$ values well
   below our detection threshold. 
   \label{fig:time}}
\end{figure*}

We have also computed $W_{H\alpha}$ as a function of time for the higher signal-to-noise dates
2016-02-04 (KELT-3) and 2016-02-23 (GJ 436) dates. For the individual measurements, we choose to
narrow the integration width to $\pm200$ km s$^{-1}$ so as to not introduce additional uncertainty
into the already noisy individual measurements.  The $W_{H\alpha}$ timeseries is shown in
\autoref{fig:time}. For each date, the first five pre-transit spectra, which correspond to the first
five points shown in each panel, are used to construct a comparison spectrum. All individual
transmission spectra are then constructed with that comparison spectrum. We note that individual
points that appear to show a 1-2$\sigma$ signal are most likely the result of small residual
differences in sky subtractions, which tend to average out in the master spectra (see
\autoref{fig:kelt3} and \autoref{fig:gj436}). In other words, we do not believe these signals are
intrinsic to the stellar system. Uncertainties for each point are the normalized flux uncertainties
added in quadrature across the range of integration.

The absence of absorption in individual KELT-3 observations is consistent with the null detection in
the combined average spectra. Although no significant H$\alpha$ changes are detected, the observations
serve as a demonstration of the achievable precision for $V$$\sim$10 stars using a high-resolution
spectrograph on a 4-meter telescope.

Absorption is also absent from the GJ 436 timeseries. We have simulated what the expected $W_{H\alpha}$
signal would be from the extended hydrogen cloud suggested by \citet{ehren15}
and revisited by \citet{bourrier16} (B16 in \autoref{fig:time}) to include interactions of the planetary outflow with the stellar wind.
The $W_{Ly\alpha}$ values (red solid line) are calculated directly from the simulated Ly$\alpha$ profiles
of Visit 2 from \citet{bourrier16}. Note that we follow both \citet{ehren15} and \citet{bourrier16} by only integrating
the absorbed Ly$\alpha$ flux between $-120$ km s$^{-1}$ and $-40$ km s$^{-1}$.

To construct the H$\alpha$ absorption from the Ly$\alpha$ measurements, we assume a simple Gaussian
line profile shape for Ly$\alpha$ and fit the optical depth, line center, and line width to
the simulated line profile from \citet{bourrier16}. We then use this simplified line profile to
solve for the $n=1$ column density:

\begin{equation}\label{eq:colden}
N_{1} = \frac{\sqrt{\pi} e^2 f_{12} \tau_{1} \lambda_{12}}{m_e c b} 
\end{equation}

\noindent where $f_{12}$ is the Ly$\alpha$ oscillator strength, $e$ is the electron charge, $m_e$ is
the electron mass, $\lambda_{12}$ is the Ly$\alpha$ rest wavelength, $c$ is the speed of light in
vacuum, and $\tau_{1}$ is the $n=1$ optical depth at line center.  We assume various
excited-to-ground state ratios for the hydrogen atoms which gives the $n=2$ column density,
$N_{H\alpha}$, based on $N_{Ly\alpha}$ and, using the same line widths, calculate the corresponding
H$\alpha$ line profile at each time.  The model $W_{H\alpha}$ values are then calculated from the
approximated line profiles.  However, since H$\alpha$ is unaffected by interstellar absorption, we
integrate from $-200$ km s$^{-1}$ to $+200$ km s$^{-1}$ to simulate the measurements made from our
observations. We do not integrate from $-500$ to $+500$ km s$^{-1}$ since this contributes nothing
to $W_{H\alpha}$ for the model line profiles. We assume that the column density at each time is
constant across the stellar disk and that the entire disk is occulted by the hydrogen cloud
\citep[see Figure 3 of][]{ehren15}. The Gaussian approximations to the Ly$\alpha$ model profiles and
the treatment of the column density from \autoref{eq:colden} yields a median $n=1$ column density of
$N_1=1.3\times10^{13}$ cm$^{-2}$. We caution that more precise column densities, as a function of
velocity, should be retrieved from the full model output of \citet{bourrier16}. 

The $W_{H\alpha}$ model values are shown as dashed-dotted lines in \autoref{fig:time} for
excited-to-ground state ratios $n_2/n_1$=10$^{-3}$ (steel blue) and 10$^{-4}$ (dark green). Note
that the right-hand axis gives the Ly$\alpha$ model values while the left-hand axis corresponds to
$W_{H\alpha}$ values, both for the model and data. Our simulated $W_{H\alpha}$ values show that,
based on the \citet{bourrier16} model, only for optimistic ratios of $\gtrsim$10$^{-3}$ would the
cloud absorption be detectable in our observations. Including hydrogen excitation in a model similar
to that from \citet{bourrier16} would be useful in order to produce more realistic estimates of the
excited hydrogen absorption. Our first-order estimates suggest that $n_2/n_1$$<$10$^{-4}$ throughout
the neutral hydrogen cloud. This is a fairly weak constraint and could be improved with more
sensitive observations from 8-10 meter class telescopes. 

\bigskip

\section{Extended atmosphere models}
\label{sec:models}

In this section we explore spherically symmetric atmospheres as a first approximation to placing
upper limits on the size of the excited hydrogen atmosphere around each planet. Two parameters
control the absorption strength: the radial extent of the atmosphere, which determines the depth of
the line, and density, which can produce opacity broadening if the column densities are large
enough.  We take the atmospheres to be of uniform density and we present 1$\sigma$ limits in two
cases: 1. varying the density and radial extent of the atmosphere to produce a value of
$W_{H\alpha}$ detectable at the 1$\sigma$ level; 2. varying the density of a filled Hill sphere,
i.e., a fixed radial extent. In both cases, we restrict the line depth to less than the 1$\sigma$
flux limit.  In Case 1, the line depth restriction is the main determinant of the radial extent.
The filled Hill sphere is an extreme case for KELT-3 b, which has a fairly high surface gravity (see
\autoref{tab:tab3}), but GJ 436 b is known to be experiencing significant mass loss and overflowing
its Hill sphere \citep{ehren15}.  The choice of uniform density is motivated by the findings of
\citet{christie} who calculated H$\alpha$ absorption in the atmosphere of HD 189733 b. 

The model consists of a uniform density atmosphere surrounding the planet at mid-transit. Planetary
and stellar parameters used in the model are given in \autoref{tab:tab3}. All parameters for the
KELT-3 system are taken from \citet{pepper13}. The GJ 436 transit data are taken from
\citet{knutson14} and the planetary mass and stellar parameter are taken from \citet{torres08}. We
fill a 3D grid with material at the specified density and extinct the intensity from the star that
is obscured by those grid points. The line profile is approximated as a Doppler--broadened delta
function \citep{draine} with an intrinsic width of $\Delta$$v=4.0$ km s$^{-1}$. This value is chosen
based on the line width necessary to reproduce the in-transit H$\alpha$ transmission spectra from
\citet{cauley15,cauley16}.  For simplicity and due to fact that no signal is detected in any $S_T$
measurements, we neglect any broadening from stellar or planetary rotation. We also neglect any limb
darkening effects from the star. The flux from the unobscured portions of the stellar disk are then
added to the absorbed grid points and the final summed line profile is normalized. The line profile
is then convolved with a Gaussian of FWHM 15 km s$^{-1}$ to approximate the Hydra instrumental
profile. 

To determine the case 1 and case 2 limits, we take the flux uncertainties from the observed
transmission spectra and apply them uniformly to the model spectra. We then propagate the flux
errors in quadrature when calculating $W_{H\alpha}$. The limits are determined when the model signal
approaches the 1$\sigma$ level.  Limits are derived only for the nights of 2016--02--04 and
2016--02--23 due to the higher quality of data. For the filled Hill sphere, or Case 2, we vary the
density until the line depth approaches the 1$\sigma$ flux uncertainty. Note that in this case the
extended atmosphere is very optically thin whereas in Case 1 the lines are highly optically thick.
  
\begin{figure}  
\begin{deluxetable*}{lccccccccccc}
\tablecaption{System parameters\label{tab:tab3}}
\tablehead{\colhead{}&\colhead{$M_{pl}$}&\colhead{$R_{pl}$}&\colhead{$R_{Hill}$}&\colhead{$P_{orb}$}&\colhead{$a$}&\colhead{$T_{eq}^1$}&\colhead{$b_{imp}$}&\colhead{Transit duration}&
\colhead{$M_*$}&\colhead{$R_*$}&\colhead{$T_{eff}$}\\
\colhead{Object}&\colhead{($M_{Jup}$)}&\colhead{($R_{Jup}$)}&\colhead{($R_{pl}$)}&\colhead{(days)}&\colhead{(au)}&\colhead{(K)}&\colhead{($R_*$)}&\colhead{(hours)}&
\colhead{($M_\odot$)}&\colhead{($R_\odot$)}&\colhead{(K)}\\
\colhead{(1)}&\colhead{(2)}&\colhead{(3)}&\colhead{(4)}&\colhead{(5)}&\colhead{(6)}&\colhead{(7)}&\colhead{(8)}&\colhead{(9)}&\colhead{(10)}&\colhead{(11)}&\colhead{(12)}}
\tabletypesize{\scriptsize}
\startdata
KELT-3 & 1.48 & 1.35 & 4.59 & 2.70 & 0.041 & 1822 & 0.61 & 3.158 & 1.28 & 1.47 & 6304\\
GJ 436 & 0.07 & 0.38 & 5.83 & 2.64 & 0.029$^{2}$ & 650 & 0.85 & 0.881 & 0.45 & 0.46 & 3350\\
\enddata
\tablenotetext{1}{The equilibrium temperature $T_{eq}$ is estimated from $T_{eq}=T_{eff}\sqrt{R_*/2a}$ \citep{charb05}}
\tablenotetext{2}{We ignore GJ 436 b's eccentricity since the atmospheric models are not time-dependent.}
\end{deluxetable*}
\end{figure}

\begin{figure*}[tbh]
   \centering
   \includegraphics[scale=.75,clip,trim=18mm 40mm 15mm 40mm,angle=0]{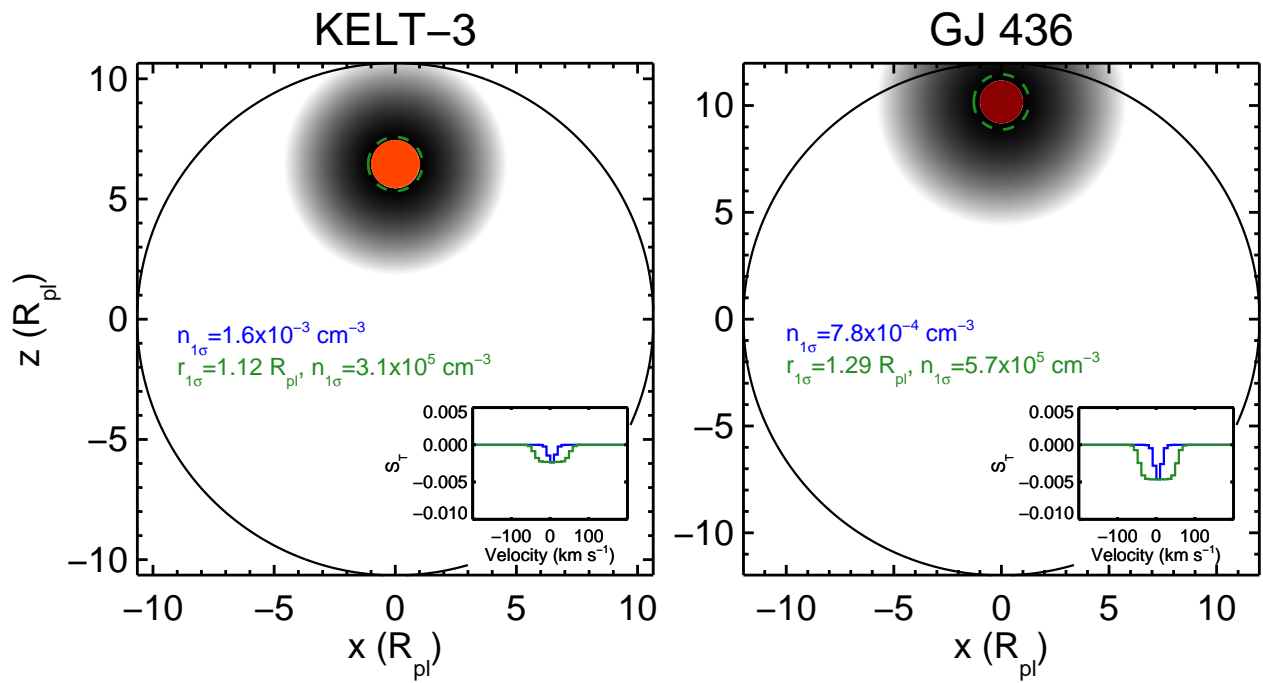} 
   \figcaption{Diagrams showing the system mid-transit geometries of full Hill spheres for both KELT-3 b (left) and
   GJ 436 b (right). The opaque planetary disks are shown in orange and red and the filled Hill spheres are shown in black. 
   The 1$\sigma$ number density limits ($n_{1\sigma}$) for the filled Hill sphere are shown in blue and the 1$\sigma$
   maximum radii ($r_{1\sigma}$) and density values for the uniform atmospheres
   are given in green. The maximum radii are also indicated with dashed green lines around the planets. The model H$\alpha$ line
   profiles for both cases are inset in the lower right corner. The velocity and flux ranges for the line profiles has been narrowed
   compared to \autoref{fig:kelt3} and \autoref{fig:gj436} in order
   to highlight the profile shapes. The broad profiles for the varying radial extent case are the result of opacity
   broadening from the very large optical depth.\label{fig:limits}}
\end{figure*}

\autoref{fig:limits} shows diagrams of the mid-transit geometry and the limits for both cases. The
model line profiles for each case are inset in the bottom-right. The radial 1$\sigma$ limits are
perhaps more interesting: it is clear that the signal-to-noise of the data, for either object, is
not high enough to rule out significant extended atmospheres of excited hydrogen. In the case of GJ
436 b, this limit extends to almost 1.3 $R_{pl}$. Given the suggested extent of the escaping neutral
hydrogen envelope, it is plausible that the base layer, which we are not sensitive enough to probe,
could produce absorption detectable with an 8-10 meter class telescope. We note that this base
thermospheric layer is independent of the outflowing material so the $n_2/n_1$ limits derived here
are not applicable. The brightness of KELT-3 would allow a similar detection if excited hydrogen
exists out to 1.1 $R_{pl}$ in similar densities to what we've modeled here. We note that the
1$\sigma$ density limits derived for the limited radial atmospheres are fairly large and that
non-negligible absorption produced by lower densities is certainly plausible.

The case of KELT-3 b can be compared to the simulations by \citet{salz16} for the hot planet WASP-77
b \citep{maxted13}, which has a slightly larger surface gravity and similar equilibrium temperature.
These fairly compact planets have neutral-to-ionized layers that form closer to the planet
\citep[see Figure 8 of ][]{salz16}, reducing the strength of the potential H$\alpha$ transmission
signal. Thus although the temperatures in the thermosphere are on the order $\sim 10^4$ K, there is
little neutral hydrogen beyond $\sim$1.2 $R_p$.

One explanation for the lack of H$\alpha$ absorption around GJ 436 b, despite the fact that large
quantities of neutral hydrogen are clearly present \citep{ehren15}, is the amount of EUV flux
received by the planet. \citet{christie} found that the H$\alpha$ absorption measured by
\citet{jensen12} required an ionization rate larger than the nominal value by a factor of $\sim$3.5.
This suggests that a significant $n=2$ population requires a large amount of EUV and XUV flux. The
estimated EUV and XUV flux from GJ 436 is lower than that of HD 189733 by $\sim$1.5 magnitudes, in
both cases \citep{salz16}. Thus although there is abundant neutral hydrogen around the planet, it is
likely almost entirely in the ground state, a conclusion that is supported by the time series
analysis in \autoref{sec:timeseries}. The low peak temperature of $\sim$5000 K in the thermosphere,
as simulated by \citet{salz16}, may also be the reason for the lack of $n=2$ hydrogen: an excitation
temperature of $\sim$14300 K is needed to produce $n_2/n_1 \sim 10^{-3}$.  However, we caution that
hot planets experience highly non-equilibrium conditions and thus cannot be assumed to have a single
steady-state temperature in their extended atmospheres. Variability in absorption signatures in
ground-state and excited hydrogen may be connected with the variable activity levels of the host
star.

\section{Summary and conclusions}
\label{sec:conclusions}

We have presented a search for excited hydrogen absorption in H$\alpha$ for the exoplanets KELT-3 b
and GJ 436 b using the high-resolution Bench Spectrograph with Hydra on the WIYN 3.5-meter
telescope. We report no detections of an extended atmosphere for either transit of either target. We
also find no evidence for absorption by unbound circumplanetary material. We derive limits on the
radial extent of uniform density atmospheres of $n=2$ hydrogen and find that both planets may host
non-negligible atmospheres that produce line profiles below the signal-to-noise threshold of our
observations. More sensitive observations with larger aperture telescopes could probe below these
limits.

Our exploratory observations demonstrate the approximate limits of high-resolution transmission
spectroscopy with 4-meter class telescopes. Future missions such as the \textit{Transiting Exoplanet
Survey Satellite} (\textit{TESS}) will find many hot planets orbiting nearby bright stars. We also
expect \textit{TESS} to find a significant number of hot planets transiting \textit{active} stars,
which will be the preferred targets for detecting H$\alpha$ absorption based on the results here and
in \citet{jensen12} and \citet{cauley15,cauley16}.  These bright systems will also be ideal targets
for 4-meter class transmission spectroscopy, allowing more efficient atmosphere detections and
freeing up larger telescopes for observations of fainter targets. New methods for detecting extended
atmospheres are needed once \textit{HST} is retired and before the next space-based UV observatory
is commissioned. Searching for Balmer line absorption with ground-based optical spectrographs offers
a possible solution. 

\bigskip

{\bf Acknowledgments:} The authors thank the referee for their comments, which helped improve this
manuscript. Data presented herein were obtained at the WIYN Observatory from telescope time
allocated to NN-EXPLORE through the scientific partnership of the National Aeronautics and Space
Administration, the National Science Foundation, and the National Optical Astronomy Observatory.
This work was supported by a NASA WIYN PI Data Award, administered by the NASA Exoplanet Science
Institute. This work was also completed with support from the National Science Foundation through
Astronomy and Astrophysics Research Grant AST-1313268 (PI: S.R.). A. G. J. is supported by NASA
Exoplanet Research Program grant 14-XRP142-0090 to the University of Nebraska-Kearney.  This work
has made use of NASA's Astrophysics Data System. The authors would like to acknowledge Marla Geha
for supplying the basic Hydra IDL reduction routines. We are also grateful to Vincent Bourrier and
David Ehrenreich for providing the GJ 436 Ly$\alpha$ model data.

\end{document}